\documentclass[iop,apjl]{emulateapj}\usepackage{apjfonts}

\shorttitle{Expanded VLA and 36.2 GHz Methanol Masers in Sgr A} 
\shortauthors{Sjouwerman, Pihlstr\"om \& Fish}

\begin{document}

\title{Expanded VLA Detection of 36.2 GHz Class\,I Methanol Masers in Sagittarius\,A}

\author{Lor\'ant~O.~Sjouwerman$^1$, Ylva~M.~Pihlstr\"om$^{2\footnote{\rm Also an Adjunct Astronomer at the National Radio Astronomy Observatory}}$, Vincent~L.~Fish$^3$} \affil{$^1$ National Radio Astronomy Observatory, P.O.~Box O, Socorro, NM 87801} \affil{$^2$ Department of Physics and Astronomy, University of New Mexico, 800 Yale Boulevard NE, Albuquerque NM 87131} \affil{$^3$ Massachusetts Institute of Technology, Haystack Observatory, Route 40, Westford, MA 01886}


\email{\ \\ lsjouwerman@nrao.edu, ylva@unm.edu, vfish@haystack.mit.edu}

\begin{abstract}
We report on the interferometric detection of 36.2 GHz Class\,I
methanol emission with the new 27--40 GHz Ka band receivers available
on the Expanded Very Large Array (EVLA). The brightness temperatures
of the interferometric 36 GHz detections unambiguously indicate for the first time that
the emission is maser emission. The
36 GHz methanol masers are not co-spatial with 1720 MHz OH masers, indicating
that the two species trace different shocks. The 36 GHz
and 44 GHz methanol masers, which both are collisionally pumped, do
not necessarily co-exist and may trace different
methanol gas. The methanol masers seem correlated with NH$_3$(3,3) density 
peaks. We favor an explanation in which the 36 GHz Class\,I
methanol masers outline regions of cloud-cloud collisions, perhaps
just before the onset of the formation of individual massive stars.

The transition of the Very Large Array (VLA) to the EVLA is
well under way, and these detections demonstrate the bright future of
this completely renewed instrument.
\end{abstract}


\section{Introduction}\label{intro}

Methanol (CH$_3$OH) masers are typically classified as Class\,I and
Class\,II methanol masers. 
The Class\,II methanol masers may be pumped by strong far infrared
radiation and are related to star forming regions. Class\,I methanol
masers are probably pumped by collisions and are found in regions
where outflows or clouds collide.
This distinction is based on
observations using sufficient angular resolution, where Class\,II
masers are found to be co-spatial ($\sim$~1\arcsec) with ultracompact
HII regions whereas Class\,I masers are near (few arcseconds), but not
co-spatial with HII regions \citep[e.g.,][]{ment91,liwi96}.

Bright 36 GHz methanol emission has been found in many regions such as
the Galactic center source Sgr A, Sgr B2, OMC1 and 2, and
other Galactic star forming regions. It is usually found in sources
with other Class\,I masers, such as the 44 and 95 GHz lines
\citep[e.g.,][]{hasc90,ment91,muel04,fish07}. Due to the fact that all
previous observations at 36 GHz have been taken with single-dish
telescopes, it has heretofore not been unambiguously established that
the 36 GHz methanol emission is due to maser action
\citep{ment91,prat08}

In Sgr A, the brightest 36 GHz line emission is found in the
northeastern part of the Sgr A East supernova remnant (SNR), located
near the region where the SNR interacts with the M$-$0.02$-$0.07
molecular cloud 
\citep[intensity $\gg$ 1 Jy\,beam$^{-1}$ using a single dish, e.g.,][]{mori85,haba89,szcz89,hasc89,liwi96}.  Recently,
\citet{sjpi08} detected several 1720 MHz OH masers in the same region (Fig.~\ref{f1}).
These OH masers are collisionally pumped and are considered typical
signposts of shock excited material. Theoretical modeling indicates a similar range of
physical properties (T$_{\rm K}$ $\sim$ 80-100 K, $n \sim$
10$^4$-10$^5$ cm$^{-3}$) for 36 GHz Class\,I methanol
and 1720 MHz OH maser excitation
\citep{mori85,prat08,pihl08,ment09}. That is, if there is methanol
abundance near a 1720 MHz OH maser, there may be 36 GHz Class\,I
methanol masers.  This is supported by the fact that locations of the
northeastern 1720 MHz OH masers in Sgr\,A are consistent with the core
of the 36 GHz methanol emission.

The new 27--40 GHz (Ka band) receivers recently commissioned on the
(Expanded) Very Large Array \citep[EVLA/VLA,][]{ulve06}
enable interferometric high angular resolution
observations of sources hosting 36 GHz masers.  Here we report
on the first interferometric observation of the Sgr\,A region in the
36 GHz line of methanol with the EVLA to investigate whether this
emission may be due to maser activity and, if so, whether it is
co-spatial with the shock-excited 1720 MHz OH masers. We report on
initial results confirming the maser nature of the 36 GHz methanol
emission and identifying their location relative to
the 1720 MHz OH masers. A future paper will
provide a more complete analysis of our observations.

\section{Observations}\label{observations}

\begin{figure} \begin{center}
\includegraphics[height=0.75\columnwidth, angle=-0]{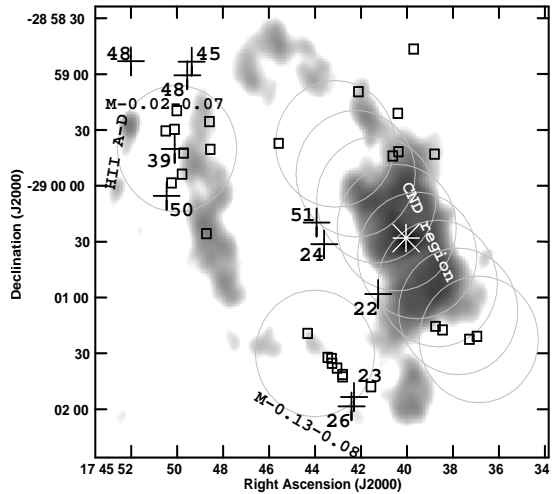}\end{center}
\caption{Gray scale outline of Sgr A in 1.7 GHz radio continuum with
  1720 MHz OH masers (squares; see also \citealt{sjpi08}). The newly detected 36 GHz
  methanol masers are located at the plus-signs and
  labeled with their LSR velocities. The closest angular separation
  between any methanol and OH maser is 6.5\arcsec\ (0.25
  parsec). Sgr\,A* is located in the dark CND region to the
  west (white star). The large gray circles show the 68\arcsec\ 
  diameter primary beams; the core of M$-$0.02$-$0.07 is covered by 
  the easternmost beam.
\label{f1}
}
\end{figure}

The VLA was used at 36 GHz ($\lambda$ = 8.3 mm) in 2009 March and
April, after 9 EVLA type Ka band receivers were deployed. The
receivers were placed within the B array configuration to optimize the
sampling in the \emph{(u,v)}-plane, with the longest baseline
measuring $\sim$7.5 km (870 k$\lambda$). This ``limited B array'' used
baselines of 70 k$\lambda$ and up and resulted in a synthesized beam
of about 200-400 mas in a primary beam (field of view) of about
68\arcsec\ diameter. However, in March only 7 antennas yielded usable
data.  Because the new receivers have a new frequency tuning
system, a new observation preparation tool (OPT) was used. This new
OPT software is a web based application that will completely replace
the tools currently used to prepare observations for the VLA
(``Jobserve'').

The VLA correlator was configured to deliver Doppler tracked data at
3.3 second integration intervals for a single dual polarization IF
pair using 6.25 MHz total bandwidth ($\sim$50 km\,s$^{-1}$). This bandwidth
was divided in 63 channels at 0.8 km\,s$^{-1}$ (97.7 kHz) separation and tuned
to the $J\,=\,4_{-1}\,\rightarrow\,3_{0}E$ rotational transition of
methanol at 36\,169.265 MHz \citep{muel04}. The amplitude scale was
adopted using the standard spectrum of 3C286 \citep{baar77}; pointing, bandpass and complex gain
calibration was performed using NRAO530 and Sgr\,A*.

In March, 8 different position/velocity settings covered the
circumnuclear disk (CND) around Sgr\,A* over two arcminutes and
$\pm$150 km\,s$^{-1}$ in the reference frame of the Local Standard of Rest
(LSR). Each was observed for $\sim$15 minutes. Additionally,
$\sim$3 minute on source pointings were included on two regions where many
1720 MHz OH masers are located. One of them pointed at 
the core of M$-$0.02$-$0.07 toward the east
(Fig.~\ref{f1}). In April selected fields were reobserved with 3--4
minute on source integration times and with shifted Doppler velocity
settings to confirm potential detections from the first run.

\section{Results}\label{results}

Table~\ref{t1} and Figs.~\ref{f1} and \ref{f2} summarize our results
for selected 36 GHz methanol masers (Sect.~\ref{s4.1}). These masers
are all redetected in our different observations and are either the
brightest members of clusters of masers found near the northeastern
1720 MHz OH masers or individual (weaker) masers near the
southeastern 1720 MHz OH masers and the CND. Our complete list of tens
of detections will be given in a later manuscript.

The data reduction, performed with the AIPS package, was complicated
by the strong methanol lines that were present in all pointings except
for the highest absolute velocities (at $|{\rm V}_{\rm LSR}| > $ 80
km\,s$^{-1}$). We produced image cubes that were cleaned well beyond
(2.5 times) the primary beam, since significant emission was detected
beyond the primary beam in several pointings (Fig.~\ref{f1}).
We derived
approximate flux densities of 1.7, 3.6($\pm$0.1) and 1.1($\pm$0.1) Jy
for our calibrator sources 3C286, NRAO530 and Sgr\,A*, respectively.
Typical values for the RMS image noise in a line-free channel were
10--12 mJy\,beam$^{-1}$.

Table~\ref{t1} lists selected maser detections with the position and
velocity of the channel of peak emission, the span of
velocities over which emission was detected, 
the integrated line flux, and a lower limit
of the brightness temperature (derived from the flux density at 
the line peak). The masers on the eastern side of Sgr\,A were
self-calibrated. The masers near the CND are outside our
primary beam and self-calibrated using Sgr\,A*.  But because we have
confirmed their positions within 0.3\arcsec, based on
redetections of these masers in the April data,
we can be confident that the positions are accurate enough to place
the masers among the other phenomena in the Sgr\,A region, such as the
1720 MHz OH masers in Fig.~\ref{f1} and the HCN and NH$_3$(3,3) emission in
Fig.~\ref{f2}. 

\begin{table}
\begin{center}\caption{Results on selected detections
}\label{t1}\begin{tabular}{rccccrc}
\multicolumn{3}{r}{Right Ascension \& Declination} &V$_{\rm LSR}$ &$\Delta$V &Flux &T$_{\rm B}$\\
&\multicolumn{2}{c}{(J2000)} &\multicolumn{2}{c}{(km\,s$^{-1}$)} &$^\dagger$ &(K)  \\
\hline\hline
\noalign{\smallskip}
 1$\phantom{^\ast}$&17 45 41.25&$-$29 00 58.1&+21.5&1.6&  0.95&$> 1.1\cdot 10^4$ \\ 
 2$\phantom{^\ast}$&17 45 42.30&$-$29 01 53.4&+23.1&0.8&  0.67&$> 1.4\cdot 10^4$ \\ 
 3$\phantom{^\ast}$&17 45 42.41&$-$29 01 58.5&+25.6&1.6&  0.76&$> 9.7\cdot 10^3$ \\ 
 4$\phantom{^\ast}$&17 45 43.60&$-$29 00 31.4&+24.0&1.6&  3.86&$> 6.7\cdot 10^4$ \\ 
 5$\phantom{^\ast}$&17 45 43.93&$-$29 00 19.7&+50.7&1.6&  0.52&$> 9.1\cdot 10^3$ \\ 
 6$^\ast$&17 45 49.36&$-$28 58 53.3&+45.0&7.3& 154  &$> 8.8\cdot 10^5$ \\ 
 7$^\ast$&17 45 49.55&$-$28 59 00.6&+48.2&5.7& 62.4 &$> 4.6\cdot 10^5$ \\ 
 8$^\ast$&17 45 50.10&$-$28 59 40.2&+39.3&4.9& 215  &$> 2.3\cdot 10^6$ \\ 
 9$^\ast$&17 45 50.45&$-$29 00 05.4&+49.9&3.2& 60.4 &$> 6.8\cdot 10^5$ \\ 
10$^\ast$&17 45 52.00&$-$28 58 53.1&+48.2&1.6& 35.1 &$> 6.1\cdot 10^5$ \\ 
\noalign{\smallskip}
\hline
\multicolumn{7}{l}{$^\dagger$: Line flux in Jy$\cdot$km\,s$^{-1}$}\\
\multicolumn{7}{l}{$^\ast$: Brightest component of a cluster of detections}\\
\end{tabular}\end{center}
\end{table}

An absolute flux scale for Ka band has not yet been formalized, but the flux
density measured for Sgr\,A* (1.1 Jy) is
consistent with the expected flux at this wavelength,
thus lending confidence in our flux scale.  We have modified our fluxes
 in Table~\ref{t1} for
the yet-undetermined primary beam attenuation of the EVLA antennas at Ka
band using a simple Gaussian beam pattern with the same FWHM as our
primary beam.

\section{Discussion}\label{discussion}

\subsection{Some 36 GHz Methanol Emission is \emph{Maser} Emission}
\label{s4.1}

Until recently, 36 GHz methanol emission was solely observed with
single dish instruments, with typical spectra showing narrow peaked
emission, sometimes strong, on top of a broader wide base
\citep[e.g.][]{mori85,haba89,szcz89,szcz91,liwi96}. Because of the
limited angular resolution, 36 GHz methanol modeling has not been
heavily pursued \citep[however see, e.g.,][for individual modeling of
the 36 GHz methanol line]{mori85,crag92,liwi96,ment09}. Most authors
have convincingly argued that the spectra indicate a broad thermal
emission component with strong, narrow maser emission components
superimposed.
Due to the large beam sizes of the single dish
telescopes used, 
the lower limits these authors have placed on the brightness
temperatures of these narrow components has not been high enough to prove
unambiguously that the emission mechanism is maser, not thermal. \citet{liwi96} present a nice discussion, including a
counter example (their Sect.~4.1). The interferometric point-like
detection of the 36 GHz methanol emission in this work, utilizing a
much smaller synthesized beam size than with the single dish observations
previously, has clearly put lower limits on the brightness
temperature of several masers above 10$^5$ K (Table~\ref{t1}). That is, the 36 GHz
methanol emission must be \emph{maser} emission, as suspected by
previous authors.

\subsection{1720 MHz OH versus 36 GHz Methanol Maser Emission}
\label{s4.2}

Both (36 GHz) Class\,I methanol and 1720 MHz OH maser emission is
generally considered to be tracing shock interaction regions.
\citet{szcz89,szcz91} and \citet{liwi96} have published mosaics of 36
GHz methanol emission in the Sgr\,A region using the single dishes of
Haystack and Effelsberg (with 40\arcsec\ and 25\arcsec\ angular
resolution, respectively), consistent with each other and this work. The direction of
the CND was only covered by \citet{szcz91}, although they did not use
a bandwidth wide enough to observe the full range of LSR velocities
($\pm150$ km\,s$^{-1}$) present in the CND. Their methanol results are
consistent with the parsec scale location and kinematics of the cores
of the giant molecular clouds M$-$0.02$-$0.07 and M$-$0.13$-$0.08
(also known as the $+$50 km\,s$^{-1}$ and $+$20 km\,s$^{-1}$ clouds 
respectively; \citealt{zylk90}) as
determined in many other molecular lines and infrared dust emission.
The broad line profile of the methanol emission was suggested to
originate from material shocked and compressed by the impact of nearby
SNRs (Sgr\,A-East and G359.02$-$0.09 respectively), whereas the narrow
profiles were postulated to be due to maser emission
(Sect.~\ref{s4.1}).
The location of shock excited 1720 MHz OH masers is also consistent 
with this picture \citep{sjpi08}. That is, the OH masers and methanol emission
both originate in the molecular material that is being shocked by the
expansion of SNR material.

We detect tens of 36 GHz methanol masers toward the core of
M$-$0.02$-$0.07, of which the brightest in each cluster of detections
are shown in Table~\ref{t1} and Fig.~\ref{f1}.  The very brightest
maser ($\gg$ 100 Jy) was probably previously detected as the ``spike''
of emission in the ``Grid C'' spectrum of \citet{szcz89}. The methanol
maser positions are well separated ($\sim$ 15--30\arcsec, 0.5--1.2
parsec) from, and to the west of the HII regions A--D. They are placed
amongst the 1720 MHz masers (Fig.~\ref{f1}) with velocities consistent
with the molecular cloud. This location, away from A--D toward the west,
is also consistent with the location of the ``Grid C'' spectrum of
\citet[][their Fig.~2]{szcz89}.  We did not detect a thermal component
of the molecular cloud with the long baselines used, but the masers
themselves do not explain all emission seen by \citet{szcz89,szcz91}
and \citet{liwi96}. We deduce that the \emph{thermal} emission detected by
the single dish observations is consistent with originating in the
molecular cloud core and the \emph{maser} emission is consistent with
an origin in the cloud core or
west of the cloud core near the interaction region with the SNR
expansion.

The velocities of the methanol and OH masers in the northeastern part
of Sgr\,A differ by 5--30 km\,s$^{-1}$, with the methanol velocities most
compatible with the kinematics of the core. This suggests that the
methanol masers are more closely related to the cloud core whereas the
OH masers are likely outlining post-shock gas from the interaction of
the SNR with the molecular cloud. We note the striking alignment of
methanol masers with the northeastern part of the continuum rim of
the SNR, where the northeastern 1720 MHz OH masers are more scattered.
We also note the striking alignment of three methanol masers
with the outer contours of the HCN emission of the CND (see
Sect.~\ref{s4.4}), and the alignment of methanol masers with the
southeastern OH masers. However, as the minimum angular separation
between a methanol-OH maser pair in any of our detections is
6.5\arcsec\ (0.25 parsec), we can conclude that 1720 MHz OH and 36 GHz
methanol masers are not co-spatial (within $\sim$~1\arcsec). The
general difference in velocity between OH and methanol masers suggests
that they trace different shocks, or different gaseous components of a
shocked region if both maser transitions are due to a shared origin.
Perhaps this indicates that the methanol is located deeper in the
cloud, where the line-of-sight LSR velocity is not as much disturbed
by the SNR shock front as the region where the OH masers originate,
and perhaps requiring less energetic shock excitation than OH.  The
location and physical properties of the gas giving rise to the 36 GHz
methanol masers \citep[T $<$ 100 K and $n \sim 10^4-10^5$
cm$^{-3}$:][]{mori85,crag92,liwi96,ment09} also somewhat differ from
the gas traced by 1720 MHz masers \citep[T $\sim$ 75 K and $n \sim
10^5$ cm$^{-3}$:][]{pihl08}.

\begin{figure} \begin{center}
\includegraphics[height=0.75\columnwidth, angle=-0]{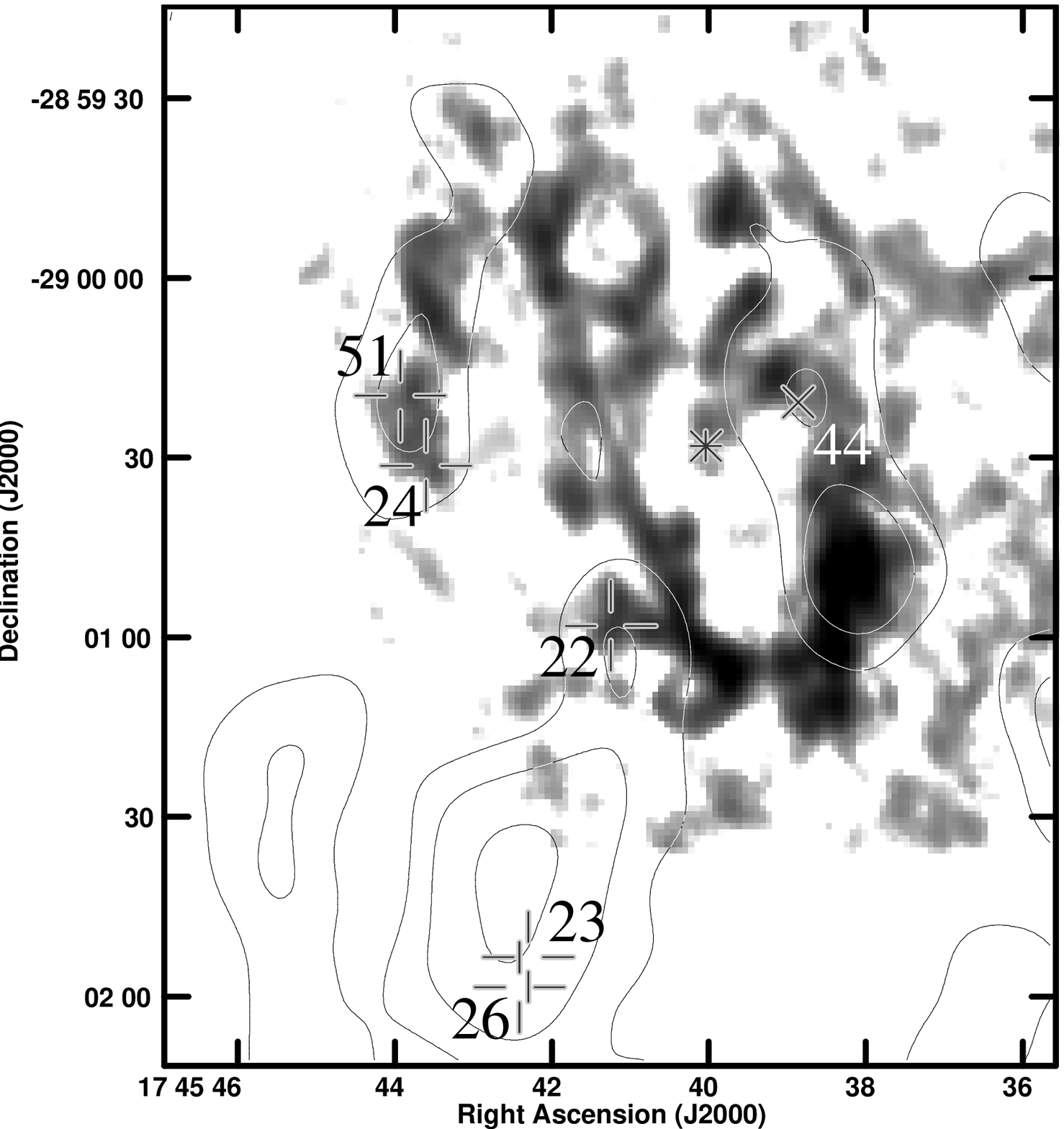}\end{center}
\caption{Integrated intensity contours of NH$_3$(3,3) emission and integrated intensity gray scale of HCN(4-3) emission outlining the
  ring-like CND \citep[see also][]{mcga01,mont09}. The newly detected 36
  GHz methanol masers near the CND (plus-signs) are labeled with their LSR velocities, as is the 44 GHz methanol maser ``V'' (cross) \citep{yuse08}. The star
locates Sgr\,A* for reference. The HCN(4-3) image only partly covers the frame.
\label{f2}
}
\end{figure}

We note that no bright 36 GHz masers were detected in the tangent
points of the CND, at V$_{\rm LSR} \sim \pm$130 km\,s$^{-1}$, where 1720 MHz
OH masers indicate that energetic shocks are present. Perhaps these
regions are too warm,  too dense, or devoid of methanol. Alternatively, these shocks 
may be too strong for 36 GHz methanol masers to exist.

\subsection{Molecular clumps and Methanol Maser Emission}
\label{s4.3}

Three 36 GHz methanol masers were detected near and (south)east
of the CND (Figs.~\ref{f1} and \ref{f2}). These masers are located just
beyond the primary beam, but can be observed in the field of
our phase calibrator (Sgr\,A*) as well as in two other observed
fields.  With the single dish Green Bank telescope ($\sim$ 15\arcsec\ 
angular resolution), \citet{yuse08} place two
bright 44 GHz Class\,I methanol masers in the same region (see
Sect.~\ref{s4.4}), located westward and consistent with the velocity
of the HCN(1-0) ``clumps'' ``F'' and ``G'' \citep{chri05}, which are 
the HCN(4-3) ``clumps'' ``CC'' and ``BB'' \citep{mont09}. Two of the 36
GHz masers are located near clump ``G''/``BB'' and another is located 
eastward of HCN(1-0) clump
``J''/``K'' \citep{chri05}, in HCN(4-3) clump ``X'' \citep{mont09}. They all have a 
velocity distinct from nearby
HCN emission in the CND, although HCN emission is present at the maser
velocities ($\sim 50-55$ and $\sim 20-25$ km\,s$^{-1}$;
\citealt[][their Figs.\ 4 and 7]{mont09}). It is therefore likely that
the methanol masers and the HCN clumps, located eastward and with a 
velocity deviating from the CND, are related.

It is even more interesting to note that the three methanol masers near
the CND, as well as the two detections toward the south of the CND and
the masers near the core of M$-$0.02$-$0.07, all appear to be located
in NH$_3$(3,3) density clumps \citep[][Fig.~\ref{f2}]{mcga01}.
Observations of ammonia in the Sgr\,A region have revealed
finger-like protrusions of gas toward the CND
\citep[e.g.,][]{coho00,mcga01}, suggesting the
existence of streamers and/or gas extensions from the cores of
M$-$0.02$-$0.07 (westward) and M$-$0.13$-$0.08 (northward) in the
direction of, and possibly feeding, the CND. This gas could have been
``loosened'' by the SNR impacting in the clouds. The three 36 GHz methanol
masers near the CND at the HCN clumps ``G''/``BB'' and ``X'' exactly lie 
on top of NH$_3$(3,3) emission peaks which supposedly outline where
gas is impacting into the outer
regions of the CND \citep[e.g.,][]{mcga01}.

\subsection{Class\,I 44 GHz and 36 GHz Methanol Maser Emission}
\label{s4.4}

\citet{yuse08} detected 44 GHz methanol maser emission in five regions
toward Sgr\,A with the single dish Green Bank telescope with an
angular resolution of about 15\arcsec. These authors confirmed two
detections with the VLA, but do not detail the interferometric
data\footnote{The VLA observations date from after their Letter was
first submitted.}. The single dish observations attributed  two of their
masers to the HII regions ``A'' and ``D''. We believe
their physical association with the
HII regions is unlikely, due both to the sparse angular resolution of
their observations and the fact that the HII regions are
are in the foreground \citep[e.g.][]{karl03}. With our new data, it seems likely
that these two 44 GHz masers are closely related to 36 GHz masers found in
this region.

Two 44 GHz masers, placed directly east of the CND by \citet{yuse08},
are near two 36 GHz masers detected in this work. The
northern most of these 44 GHz masers is positioned near a water maser and may
indicate an early stage of massive star formation or,
alternatively, be due to an
evolved star \citep{yuse08}. 
However, within their positional accuracy it is possible that
the brightest emission in this beam could arise from the
brighter southern 44 GHz maser at the same velocity \citep[][see their
Fig.~1]{yuse08}. This bright feature was not confirmed in their VLA
observations. The southern 44 GHz maser has a position and velocity
consistent with the 36 GHz maser at V$_{\rm LSR} =$ 51 km\,s$^{-1}$.
We determine the angular separation between the 36 and
44 GHz masers to be about 2\arcsec\ using
their Fig.~4.  

No bright 44 GHz masers appear to
have been detected at the velocity of the other two 36 GHz masers at
V$_{\rm LSR} =$ 22 and 24 km\,s$^{-1}$. In addition, no 36 GHz emission is
detected near the position of the weakest 44 GHz maser, northwest of
Sgr\,A* \citep[at V$_{\rm LSR} \sim$ 44 km\,s$^{-1}$ near HCN ``V'' or ``H'';][]{chri05,mont09,yuse08}. This 44 GHz 
maser, like all 36 GHz and probably all other 44 GHz masers, is also 
located in a NH$_3$(3,3) density clump \citep[][Fig.~\ref{f2}]{mcga01}.

We thus find 36 GHz masers probably co-spatial with 44 GHz masers as
well as isolated 36 GHz and isolated 44 GHz masers. Whether observations of
any combination of these masers \citep[e.g.][]{prat08} is direct
evidence for star formation or whether these Class\,I masers indicate
different temperatures and densities in shocked gas can now be
investigated at high angular resolution.  \citet{yuse08}
suggest that the 44 GHz masers near the CND indicate the onset of
massive star formation, but our complementary 36 GHz data suggest that
the picture might be less clear. The Class\,I methanol masers and the
suggestions of gas feeding the CND from the molecular cloud cores
\citep[e.g.,][and references therein]{szcz91,coho00} may perhaps
better be explained by cloud-cloud interactions creating shock waves
at those locations. Perhaps an arcsecond angular resolution
interferometric survey of 36 GHz (and 44 GHz) methanol in the whole Sgr\,A
region and a VLBI proper motion study of these masers will enlighten
this issue.

\


We thank Brian Truitt of the EVLA SSS group for swiftly resolving
logistical issues with submitting the OPT schedules to the dynamic
scheduler. The integrated intensity HCN(4-3) and NH$_3$(3,3) 
images (Fig.~\ref{f2}) were kindly provided by
Mar{\'{\i}}a Montero-Casta{\~n}o. The (Expanded) Very Large Array is operated by
the National Radio Astronomy Observatory, which is a facility of the
National Science Foundation operated under cooperative agreement by
Associated Universities, Inc.

{\it Facilities:} \facility{EVLA (), VLA ()}

\end{document}